\documentclass[prl,10pt,showpacs,amsmath,twocolumn,floatfix]{revtex4}
\usepackage{amsmath}
\usepackage{amsfonts}
\usepackage{graphicx}

\usepackage{mathptmx}

\newcommand{\ket}[1]{|#1\rangle}
\newcommand{\bra}[1]{\langle#1|}

\newcommand{\avg}[1]{\langle#1\rangle}

\newcommand{\bk}[1]{\left(#1\right)}
\newcommand{\Bk}[1]{\left[#1\right]}

\newcommand{\trace}{\operatorname{Tr}}
\begin{document}
\title{Quantum Imaging beyond the Diffraction Limit
by Optical Centroid Measurements}

\author{Mankei Tsang}

\email{mankei@mit.edu}


\affiliation{Research Laboratory of Electronics, Massachusetts
  Institute of Technology, Cambridge, Massachusetts 02139, USA}




\date{\today}

\begin{abstract}
  I propose a quantum imaging method that can beat the Rayleigh-Abbe
  diffraction limit and achieve de Broglie resolution without
  requiring a multi-photon absorber or coincidence detection. Using
  the same non-classical states of light as those for quantum
  lithography, the proposed method requires only optical intensity
  measurements, followed by image post-processing, to produce the same
  complex quantum interference patterns as those in quantum
  lithography. The method is expected to be experimentally realizable
  using current technology.
\end{abstract} 

\maketitle
\noindent
It has been suggested that an ensemble of photons, under appropriate
measurements, can be regarded as a single quantum object with a much
smaller effective wavelength, called the photonic de Broglie wavlength
\cite{jacobson}. The de Broglie wavelength of $N$ photons, each with
classical wavelength $\lambda$, can be as small as $\lambda/N$.  It is
especially desirable for imaging applications to take advantage of the
small de Broglie length scale, since the resolution of classical
optical imaging is limited by the size of $\lambda$, according to the
Rayleigh-Abbe diffraction limit \cite{born}. A seminal paper by Boto
\textit{et al.}\ suggests that the $N$-photon absorption patterns of
certain non-classical $N$-photon states exhibit de Broglie resolution
\cite{boto}. Subsequent work by various researchers on quantum imaging
has sought to improve upon the original ``quantum lithography''
technique \cite{agarwal,bjork,dangelo,thiel,giovannetti}, yet all of
these proposals still require a multi-photon absorber as the detector
\cite{agarwal,bjork,dangelo,giovannetti} or coincidence detection
\cite{thiel}. The low efficiency of $N$-photon absorption and
coincidence detection, especially for large $N$, is a significant
obstacle to the use of quantum technology in real-world imaging
applications.

In this Letter, I propose a quantum imaging method that achieves de
Broglie resolution without requiring a multi-photon absorber or
coincidence detection. I shall show how the same images as those in
quantum lithography can be reconstructed simply by intensity
measurements, followed by image post-processing. Given the
availability of high-efficiency single-photon detectors, the proposed
method has the potential to beat the diffraction limit by a large
amount using current technology.

A different way of achieving de Broglie resolution is via optical-beam
displacement measurements, using a split detector or homodyne
detection, as investigated by Fabre and co-workers \cite{fabre}. The
standard quantum limit to the displacement uncertainty is on the order
of $\Delta x/\sqrt{N}$, where $\Delta x$ is the classical beam width,
while the ``Heisenberg'' limit is on the order of $\Delta x/N$
\cite{fabre}. The quantum imaging theory I propose here can be
considered as a generalization of Fabre \textit{et al.}'s results, and
therefore presents a unified view of different quantum strategies that
take advantage of the photonic de Broglie wavelength. I show that the
displacement measurements can not only be considered as a Gaussian
single-parameter estimation technique, but \emph{repeated}
measurements of the parameter can also yield complex quantum
interference patterns identical to those in quantum lithography. The
proposed method thus provides an accessible way of detecting
non-classical and non-Gaussian signatures of multi-mode quantum
optical states, and is expected to become an important tool in the
study of quantum optics and quantum physics in general.

I shall first briefly review the configuration-space theory of quantum
lithography \cite{tsang_pra}. For simplicity, I consider only
free-space photons observed at the image plane in one transverse
dimension, $x$, in the paraxial regime. Generalization to two
transverse dimensions and the non-paraxial regime is possible
\cite{tsang_pra,tsang_prl}, but does not add much insight and the
results should remain qualitatively the same. Let $\hat{a}(k)$ and
$\hat{a}^\dagger(k)$ be the photon annihilation and creation
operators, respectively, in the transverse-momentum space, with the
bosonic commutation relation $[\hat{a}(k),\hat{a}^\dagger(k')] =
\delta(k-k')$.  First consider a pure $N$-photon quantum state
$\ket{N}$.  The multi-photon momentum eigenket is $\ket{k_1,\dots,k_N}
\equiv (N!)^{-1/2}
\hat{a}^\dagger(k_1)\dots\hat{a}^\dagger(k_N)\ket{0}$.  The momentum
wavefunction representation of $\ket{N}$ is then
\begin{align}
\phi(k_1,\dots,k_N) &\equiv \avg{k_1,\dots,k_N|N}.
\end{align}
The Rayleigh-Abbe diffraction limit \cite{born} restricts the
transverse momenta of photons to a finite bandwidth:
\begin{align}
\phi(k_1,\dots,k_N) &= 0
\textrm{ for any } |k_n| > \frac{2\pi \sin\theta}{\lambda},
\end{align}
where $\sin\theta$ is the numerical aperture of the optical system
and $\sin\theta \ll 1$ defines the paraxial regime.

\begin{figure}[htbp]
\centerline{\includegraphics[width=0.48\textwidth]{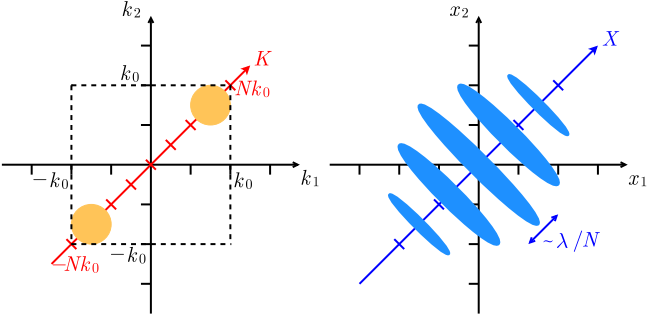}}
\caption{(Color online). Configuration-space picture of quantum
  imaging for $N = 2$. The Rayleigh-Abbe diffraction limit restricts
  the momentum wavefunction inside an $N$-dimensional box $|k_n| \le
  k_0 \equiv 2\pi\sin\theta/\lambda$. The magnitude of the total
  momentum $K = \sum_{n=1}^N k_n$ is then limited by $|K|\le
  Nk_0$. The minimum feature size of the wavefunction with respect to
  the centroid coordinate $X$ is then $\sim 1/(Nk_0) \sim \lambda/N$.}
\label{config}
\end{figure}

The spatial annihilation operator is defined as $\hat{A}(x) \equiv
(2\pi)^{-1/2}\int dk \hat{a}(k)\exp(ikx)$ and can be used to construct
$N$-photon states with definite positions $\ket{x_1,\dots,x_N} \equiv
(N!)^{-1/2}
\hat{A}^\dagger(x_1)\dots\hat{A}^\dagger(x_N)\ket{0}$.  A
multi-photon-position \emph{positive operator-valued measure} (POVM)
can be defined as
\begin{align}
\hat\Pi(x_1,\dots,x_N) =
\ket{x_1,\dots,x_N}\bra{x_1,\dots,x_N},
\end{align}
with normalization $\int dx_1\dots dx_N\hat\Pi(x_1,\dots,x_N) =
\hat{1}$. The position wavefunction is then
\begin{align}
\psi(x_1,\dots,x_N) &\equiv \avg{x_1,\dots,x_N|N},
\end{align}
which is the $N$-dimensional Fourier transform of
$\phi(k_1,\dots,k_N)$.  Changing the position variables to the
\emph{centroid} and relative-position coordinates, defined as
\begin{align}
X &\equiv \frac{1}{N}\sum_{n=1}^N x_n,
&
\xi_n &\equiv x_n - X,
\end{align}
respectively, a new POVM and a new wavefunction are obtained:
\begin{align}
\hat\Upsilon(X,\xi_1,\dots,\xi_{N-1}) &\equiv
N\hat\Pi(X+\xi_1,\dots,X+\xi_N),
\\
f(X,\xi_1,\dots,\xi_{N-1}) &\equiv \sqrt{N}\psi(X+\xi_1,\dots,X+\xi_N),
\end{align}
where $\xi_N = -\sum_{n=1}^{N-1}\xi_n$.
It is not difficult to show that the $N$-photon-absorption probability
distribution of the $N$-photon state $\hat\rho_N \equiv
\ket{N}\bra{N}$ is given by
\begin{align}
\avg{:\hat{I}^N(x):}
&=
\avg{[\hat{A}^\dagger(x)]^N[\hat{A}(x)]^N}
\propto \trace\Bk{\hat\Upsilon(x,0,\dots,0)\hat\rho_N}
\nonumber\\
&= |f(x,0,\dots,0)|^2,
\end{align}
which is proportional to the \emph{conditional} centroid probability
distribution $p_c(x)\equiv
\trace[\hat\Upsilon(x,0,\dots,0)\hat\rho_N]/\int
dx\trace[\hat\Upsilon(x,0,\dots,0)\hat\rho_N]$, given that all
relative positions are zero. The variable conjugate to the centroid
position $X$ is the total momentum $K \equiv \sum_{n=1}^N k_n$.  The
bandwidth limit on $K$ becomes $|K| \le 2\pi N\sin\theta/\lambda$, so
the minimum feature size of the centroid distribution is on the order
of $\lambda/(2N\sin\theta)$. See Fig.~\ref{config} for an illustration
of the configuration-space theory.

\begin{figure}[htbp]
\centerline{\includegraphics[width=0.4\textwidth]{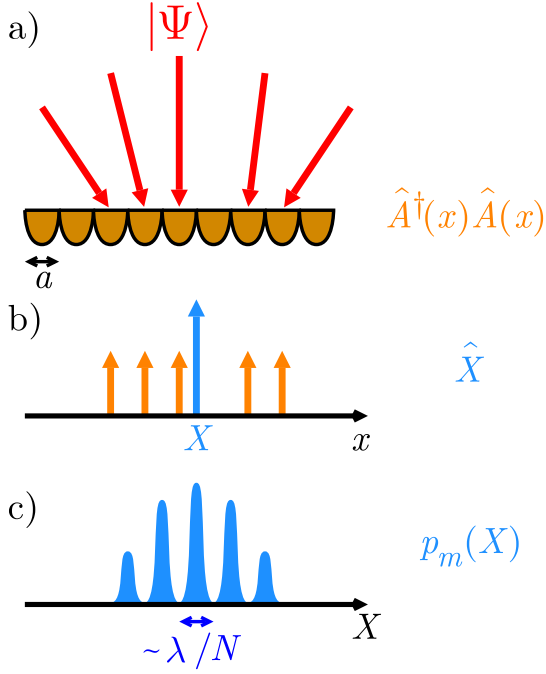}}
\caption{(Color online). a) To measure the intensity centroid
  distribution, first make an ideal spatially resolving intensity
  measurement ($\hat{A}^\dagger(x)\hat{A}(x)$).  This can be done by,
  for example, an array of photon-counting detectors, each with size
  $a \ll \lambda/(2\sin\theta)$.  b) The centroid $X$ is calculated
  from the measured intensity pattern. c) The intensity centroid
  distribution $p_m(X)$ can then be obtained after repeated
  measurements of $X$.}
\label{centroid}
\end{figure}

Multi-photon absorption is not the only way of accessing the centroid
degree of freedom. Consider the intensity centroid operator,
\begin{align}
\hat{X} &\equiv \frac{1}{N}\int dx x \hat{A}^\dagger(x)\hat{A}(x),
\end{align}
which can be measured by spatially resolving intensity measurements,
using, for example, a photon-counting detector array, followed by a
calculation of the centroid of the intensity pattern, as shown
schematically in Fig.~\ref{centroid}. The effect of finite detector
pixel size is to discretize the image plane or equivalently limit the
spatial bandwidth of the photons by virtue of the sampling theorem, so
the size of each pixel should be much smaller than
$\lambda/(2\sin\theta)$. If the pixel size is small enough, such that
the probability of more than one photon falling on each is much
smaller than one, single-photon detectors can be used instead. The
intensity operator can be rewritten as
\begin{align}
\hat{X} &= \hat{X}\hat{1}
=\hat{X} \int dx_1\dots dx_N\ket{x_1,\dots,x_N}\bra{x_1,\dots,x_N}
\nonumber\\
&=
\int dx_1\dots dx_N \bk{\frac{1}{N}\sum_{n=1}^N x_n}
\hat\Pi(x_1,\dots,x_N)
\nonumber\\
&=\int dX d\xi_1\dots d\xi_{N-1} X
\hat\Upsilon(X,\xi_1,\dots,\xi_{N-1}).
\end{align}
Thus, measurements of the intensity centroid realizes the
\emph{marginal} POVM
\begin{align}
\hat\Upsilon(X) &\equiv \int d\xi_1\dots d\xi_{N-1}
\hat\Upsilon(X,\xi_1,\dots,\xi_{N-1}),
\end{align}
and the probability distribution of the intensity centroid measurements
is 
\begin{align}
\trace\Bk{\hat\Upsilon(X)\hat\rho_N}
&= \int d\xi_{1}\dots d\xi_{N-1} |f(X,\xi_1,\dots,\xi_{N-1})|^2,
\end{align}
which is the \emph{marginal} centroid probability distribution,
hereafter denoted as $p_m(X)$.  The minimum feature size of the
marginal centroid probability distribution is similarly limited by the
bandwidth of the total momentum $|K|\le 2\pi N \sin\theta/\lambda$,
and can therefore also reach the de Broglie length scale $\sim
\lambda/N$. Unlike quantum lithography, which requires an $N$-photon
absorber for an $N$-fold resolution enhancement, intensity centroid
measurements require only photon counting and post-processing to
achieve the $N$-fold resolution enhancement for an arbitrary number of
photons.
Moreover, as the probability distribution of intensity centroid
measurements is marginal and not conditioned upon specific values of
the relative positions, the efficiency of intensity centroid
measurements is fundamentally higher than that of multi-photon
absorption or any other coincidence detection method. This is the
central result of this Letter.

In general, the conditional centroid distribution, $p_c(x)$, obtained
by multi-photon absorption, and the marginal centroid distribution,
$p_m(X)$, obtained by intensity centroid measurements, are not the
same, except for certain special quantum states. One such class of
states are the momentum-correlated or position-anti-correlated states,
of which the ``NOON'' state is a famous example
\cite{boto,tsang_pra}. These states have small uncertainties in the
relative momenta ($\kappa_n \equiv k_n - K$) and large uncertainties
in the relative positions $\xi_n$, such that the wavefunction can be
approximated as
\begin{align}
f(X,\xi_1,\dots,\xi_{N-1}) \approx g(X),
\end{align}
and $p_c(X) \approx p_m(X) \approx |g(X)|^2$. $p_c(X)$ and $p_m(X)$
become the magnitude squared of a complex wavefunction, and can
therefore exhibit the same interference patterns as those in classical
coherent imaging.  The momentum-correlated states are also the ones
that achieve de Broglie resolution, since the bandwidth of the
relative momenta $\kappa_n$ is negligible, allowing the bandwidth of
the total momentum $K$ to reach its maximum value $\sim 2\pi
N\sin\theta/\lambda$ \cite{tsang_pra}.  Such states can be created,
for example, by spontaneous parametric down conversion, and have been
experimentally demonstrated by D'Angelo \textit{et al.}\
\cite{dangelo}. The time-domain version of momentum-correlated states
has also been studied theoretically and experimentally by Giovannetti
and co-workers \cite{glm}.

Another class of quantum states with identical $p_c(X)$ and $p_m(X)$
are the ones with separable wavefunctions as follows:
\begin{align}
f(X,\xi_1,\dots,\xi_{N-1}) &= g(X)h(\xi_1,\dots,\xi_{N-1}),
\end{align}
of which the quantum Gaussian beams \cite{tsang_pra} and quantum
solitons \cite{qsoliton} are notable examples.  Classical Gaussian
beams are a special instance of quantum Gaussian beams, and both
multi-photon absorption and intensity centroid measurements of a
classical Gaussian beam produce a Gaussian spot with a width on the
order of $\lambda/(\sqrt{N}\sin\theta)$, as one would expect from a
classical theory. At the ``Heisenberg'' limit, the width of the
centroid distribution of quantum Gaussian beams is on the order of
$\lambda/(N\sin\theta)$. One can use the quantum Gaussian beam
centroid as a ``laser pointer'' to transmit and reconstruct an image
in a point-by-point fashion \cite{fabre,bjork,giovannetti}.

Next, I shall generalize the preceding results to quantum states with
indefinite photon numbers and show that the effect of indefinite
photon numbers does not significantly affect pattern formation by
intensity centroid measurements. Define a pure state as $\ket\Psi =
\sum_{N=0}^\infty C_N\ket{N}$, and denote the wavefunction of each
Fock-state component by a subscript $N$.  The $M$-photon absorption
distribution is
\begin{align}
\avg{:\hat{I}^M(x):} &\propto \sum_{N = M}^\infty
\bk{\begin{array}{c}N\\ M\end{array}}|C_N|^2
\int dx_{M+1}\dots dx_N
\nonumber\\&\quad\times
|\psi_N(\underbrace{x,\dots,x}_{M \textrm{ terms}},x_{M+1},\dots,x_N)|^2,
\end{align}
which ceases to be the centroid distribution unless $N = M$.  This
result places stringent requirements on both the generation and the
detection of non-classical states for quantum lithography.

To properly define an intensity centroid measurement, on the other
hand, first define the total photon-number operator as $\hat{N} \equiv
\int dx \hat{A}^\dagger(x)\hat{A}(x)$.  Noting that $[\hat{N},\int dx
x \hat{A}^\dagger(x)\hat{A}(x)] = 0$, the intensity centroid operator
can be more generally defined as
\begin{align}
\hat{X} &\equiv \frac{1}{\hat{N}}\int dx x \hat{A}^\dagger(x)\hat{A}(x),
\end{align}
which can be measured by first recording the
intensity pattern, discarding the results when the total photon number
is zero, and calculating the centroid of the intensity pattern.
The corresponding POVM is
\begin{align}
\hat\Upsilon(X) &\equiv \sum_{N=1}^\infty 
\int d\xi_1\dots d\xi_{N-1}\hat\Upsilon(X,\xi_1,\dots,\xi_{N-1}),
\\
\hat\Upsilon_0 &\equiv \ket{0}\bra{0}, 
\quad
\hat\Upsilon_0 + \int dX \hat\Upsilon(X) = \hat{1}.
\end{align}
The post-selected intensity centroid distribution is then
\begin{align}
p_m(X) &\propto \sum_{N=1}^\infty |C_N|^2 \int d\xi_1\dots d\xi_{N-1}
|f_N(X,\xi_1,\dots,\xi_{N-1})|^2,
\end{align}
which is the incoherent superposition of all Fock-state marginal
centroid distributions. For mixed states, the distributions are simply
the statistical average of pure-state results. If a quantum state has
a small photon-number uncertainty and most Fock components have
similar centroid distributions, the intensity centroid distribution,
being an average of the Fock-state centroid distributions, would only
be slightly smoothed by the incoherent superposition effect, whereas
for quantum lithography the effect of $N \neq M$ can be more drastic
\cite{agarwal,bjork}.

The centroid measurement can also be used to estimate the shift of an
optical beam transverse position, due to deflection by a mirror in an
atomic force microscope, for example. The results concerning the
quantum uncertainty of the beam displacement obtained by Fabre
\textit{et al.}\ \cite{fabre} can be understood using the theory
proposed here. Making the strong mean-field approximation $\hat{A}(x)
= A(x) + \Delta\hat{A}(x)$, $\hat{N}$ and $\hat{X}$ can be linearized
as $\hat{N} =N + \Delta\hat N$ and $\hat{X} = X + \Delta\hat X$, where
$\Delta\hat N \approx \int dx A^*(x) \Delta\hat{A}(x) +\textrm{H.c.}$,
\begin{align}
\Delta\hat X &\approx \frac{1}{N}\int dx x A^*(x)\Delta\hat A(x)
+\textrm{H.c.},
\end{align}
H.c.\ denotes Hermitian conjugate, and $[\Delta\hat N, \Delta\hat X]
=0$. $\Delta \hat X$ becomes a quadrature operator that can be
measured by homodyne detection with a local oscillator field $\propto
xA^*(x)$. The theory presented in this Letter is in fact more general;
it shows that repeated measurements of $\hat{X}$, without the
linearization, can not only produce a Gaussian spot, but also complex
quantum interference patterns like those in quantum lithography at the
de Broglie resolution. For position tracking applications, such as
atomic force microscopy, the proposed method can therefore accurately
determine the position of a deflecting object even if the position
significantly deviates from the mean. If the object is quantum,
intensity centroid measurements also enable the imaging of its
position wavefunction at the photonic de Broglie resolution.

Optical loss and imperfect detector efficiency are major issues for
the use of non-classical states for quantum imaging. For the specific
case of a quantum Gaussian beam, the width of $p_m(X)$ after the beam
propagates through a lossy channel can be analytically calculated
using the quantum Langevin analysis described in
Ref.~\cite{tsang_decoh}. Assuming $\avg{\hat{X}} = 0$ without loss of
generality, the result is
\begin{align}
\avg{\Delta\hat X^2}_z &=
\avg{\Delta\hat X^2}_{0} + \frac{\Delta x^2}{\eta N_z}
\bk{1-\eta e^{-\alpha z}},
\end{align}
where $\avg{\Delta\hat X^2}$ is the centroid variance, the subscript
($z$ or $0$) denotes the propagation distance, $N_z = N_{0} e^{-\alpha
  z}$ is the reduced photon number, $\Delta x^2 \equiv \avg{N^{-1}\int
  dx x \hat{A}^\dagger(x)\hat{A}(x)}$ is the classical beam width
squared, given by
\begin{align}
\Delta x^2 &=\frac{1}{4\Delta k^2}
\Bk{\frac{R_0}{N_0}+\frac{(1-1/N_0)^2}{1-1/(N_0R_0)}},
\end{align}
$\eta$ is the detector efficiency, $\alpha$ is the power loss
coefficient, and $\Delta k \sim 4\pi\sin\theta/\lambda$ is the
root-mean-square momentum bandwidth. $\Delta x$ and $\Delta k$ remain
constant in a lossy channel \cite{tsang_decoh}.  $R_0\equiv 4
N_0\Delta k^2 \avg{\Delta\hat{X}^2}_0$ is the initial normalized
centroid variance, with $R_0 = 1$ at the standard quantum limit (SQL)
for classical Gaussian beams and $R_0 = 1/N_0$ at the ``Heisenberg''
limit.  In the classical case, $R_0 = 1$, $\Delta x = 1/(2\Delta k)$,
$\avg{\Delta X^2}_z = \Delta x^2/(\eta N_z)$, and the centroid
variance remains at the SQL for the detected photon number. Close to
the ``Heisenberg'' limit, however, $\Delta x$ becomes much larger than
$1/(2\Delta k)$, and the increase in centroid variance becomes much
more sensitive to loss.  As long as $R_0 \gg 1/N$, quantum imaging by
intensity centroid measurements is fairly robust to loss.  The effect
of loss for more complex quantum images may be similarly studied using
the Langevin method, but is beyond the scope of this Letter.

Given the recent progress in the spatial engineering of biphoton
states \cite{agarwal,dangelo} and photon-counting detector technology,
the proposed method is expected to be immediately realizable using
current technology, at least for $N = 2$. For $N > 2$, it is more
challenging to create the required non-classical states of light.
Because of the post-processing requirement, the proposed method may
not be directly applicable to lithography, while the difficulty of
encoding an image in the optical centroid degree of freedom limits its
practicality in conventional imaging applications apart from image
transmission.  The proposed method nonetheless greatly simplifies the
detection of multi-mode non-classical optical states, and should be
useful, at the very least, as a diagnostic tool in the study of
quantum physics.

Discussions with Jeffrey Shapiro, Seth Lloyd, Lorenzo Maccone,
and Vittorio Giovannetti are gratefully acknowledged.  This work is
financially supported by the Keck Foundation for Extreme Quantum
Information Theory.

\end{document}